\documentclass[12pt]{article}

\usepackage{times}

\usepackage{graphicx,bm,amsmath,amsfonts,amssymb}
\newcommand{\be}{\begin{equation}}
\newcommand{\ee}{\end{equation}}
\newcommand{\beq}{\begin{eqnarray}}
\newcommand{\eeq}{\end{eqnarray}}
\newcommand{\ba}{\begin{array}}
\newcommand{\ea}{\end{array}}
\newcommand{\bea}{\begin{eqnarray}}
\newcommand{\eea}{\end{eqnarray}}
\newcommand{\ex}[1]{\mbox{e}^{#1}}
\newcommand{\vp}{\varphi}
\newcommand{\eps}{\epsilon}
\newcommand{\im}[1]{\mbox{Im}\left[#1\right]}

\newcommand{\mcal}[1]{\mathcal{#1}}

\newcommand{\g}{\gamma_{\perp}}

\newcommand{\bma}{\begin{matrix}}
\newcommand{\ema}{\end{matrix}}

\newcommand{\bx}{\bm{x}}
\newcommand{\bxp}{\bm{x}^{\prime}}

\begin{document}
\title{Strong Interactions in Multimode Random Lasers:\\ Supporting Online Material}
\author
{Hakan E. T\"ureci$^{1\ast}$, Li Ge$^{2}$, Stefan Rotter$^{2\dagger}$,  A. Douglas Stone$^2$ \\
\\
\normalsize{$^{1}$Institute of Quantum Electronics, ETH Z\"urich, 8093 Z\"urich,
   Switzerland}\\
\normalsize{$^{2}$Department of Applied Physics, P. O. Box 208284,} \\
\normalsize{Yale University, New Haven, CT 06520-8284, USA}\\
\\
\normalsize{$^\ast$To whom correspondence should be addressed;
E-mail:  tureci@phys.ethz.ch.} \\
\normalsize{$^\dagger$Presently on leave from Technische Universit\"at Wien,}\\
\normalsize{Wiedner Hauptstra\ss e 8-10/136, A-1040 Vienna, Austria, EU}
}
\date{}
\maketitle

\noindent
This preprint contains the Supporting Online Material for our paper
''Strong Interactions in Multimode Random Lasers'', Science 
{\bf 320}, 643 (2008) (also available at: arXiv:0805.4496).

\subsection*{Materials and Methods 1: Derivation and solution of the
self-consistent equations -- Details of the algorithm}

\paragraph*{Derivation of the self-consistent multimode laser equations.}
The Green function which forms the kernel of the fundamental
self-consistent equation Eq.~1 has a non-hermitian
spectral representation of the form ({\it S1})
\begin{equation}
G(\bx,\bxp ; k) =  \sum_{m=1}^{\infty}
\frac{\varphi_{m}(\bx,k)\bar{\varphi}^*_{m}(\bxp,k)}{ 2  \, k_a
(k - k_m(k))}.
\label{eqspecrep}
\end{equation}
where the CF states $\varphi_{m}(\bx,k)$ satisfy the equation
\begin{equation}
-\frac{1}{\eps(\bx)} \nabla^2 \varphi_m(\bm{x},k) = k_m^2(k) \,
\varphi_m(\bm{x},k)
\label{eqcfgenev}
\end{equation}
with the boundary condition that outside the gain region there are
only outgoing waves with wavevector $k$.  The biorthogonal partners
$\bar{\varphi}_m(\bx,k)$ satisfy the complex conjugate equation with
incoming wave boundary conditions,  hence
$\bar{\varphi}_m(\bx,k)=(\eps(\bx) \varphi_m(\bx,k))^*$.  The
biorthogonality condition on these functions is $\int_{gain} d\bx \,
\bar{\varphi}_m^*(\bx,k) \varphi_n(\bx,k) = \delta_{mn}$ with
appropriate normalization. Writing $G(\bx,\bxp ; k)$ in
Eq.~1 in the form Eq.~\ref{eqspecrep}, substituting the CF
expansion of the unknown lasing modes $\Psi_\mu (\bx) =
\sum_{m=1}^{\infty} a^\mu_m
\varphi_m(\bx,k_\mu)$, using the biorthogonality relation and
truncating to $N_{CF}$ CF states nearest the atomic line (gain
center) yields Eq.~2, with the non-linear operator
$T_{mn}^\mu = T_{mn}(k_\mu ;\bm{a}^\mu)$ where
\begin{eqnarray}
T_{mn}(k;\bm{a}^\mu) = \frac{i\g}{\g - i (k-k_a)} \frac{1}{2k_a
(k - k_m(k))} & \nonumber \\
     \times  \int  d\bxp \frac{ (1 + d_0(\bxp))\bar{\vp}^*_m (\bxp,k)
\vp_n (\bxp,k)}{\epsilon(\bxp)
(1 + \sum_\nu \Gamma_\nu |\Psi_\nu (\bxp)|^2)} &.
\label{eqTmatrix}
\end{eqnarray}
We refer to $T_{mn}(k;\bm{a}^\mu)$ also as $T_{mn}(k)$ or
$T_{mn}(k;D_0)$ in the text in order to emphasize the dependence that
is relevant in the discussion.

\paragraph*{Threshold Matrices: }
The set of non-linear equations~(2) have the following properties.
Below some finite value of the pump $D_0$ only the trivial solution exists: $\Psi_\mu
=0$, $\forall \mu$.
As $D_0$ is increased, a series of thresholds are reached at which
the number of non-trivial solutions
increases by one.  If we denote the threshold $D_{th}^{(l)}$ for
$l$-mode lasing there exist
$l$ solutions $\{\bm{a}^\mu,k_\mu \}$ ($\mu=1,\ldots,l$) for a pump
parameter $D_0$
such that  $D_{th}^{(l)}<D_0<D_{th}^{(l+1)}$.  In each of these
intervals we assume
that $l$ solutions to Eq.~2 exist and find them by a method to be
described below.

In order to find the first threshold we consider the linear operator
$T_{mn}(k;\bm{a}^\mu = 0) \equiv \bm{\mcal{T}}^{(0)}(k)$ where
\begin{equation}
\mcal{T}_{mn}^{(0)}(k) = \frac{i\g}{\g - i (k-k_a)} \frac{1}{2k_a
(k - k_m(k))}
      \int  d\bxp \frac{ (1 + d_0(\bxp))\bar{\vp}^*_m (\bxp,k)
\vp_n (\bxp,k)}{\epsilon(\bxp)}
\label{eqT0matrix}
\end{equation}
which is obtained by neglecting the $\sum_\nu \Gamma_\nu |\Psi_\nu
(\bxp)|^2$ term in the denominator of Eq.~\ref{eqTmatrix}.  The 
resulting linear equation
associated with (2) has the form
\begin{equation}
\bm{\mcal{T}}^{(0)}(k) \bm{a}^\mu = (1/D_0) \bm{a}^\mu,
\label{eqT0eval1}
\end{equation}
which has solutions (as noted in the text) when an eigenvalue of this matrix,
$\lambda^{(0)}_n(k), n=1,\ldots,N_{CF}$, is
real and has the value $1/D_0$ (i.e. $D_0 \lambda^{(0)}_n = 1$).  Since
$\bm{\mcal{T}}^{(0)}(k)$ is independent of $D_0$, so are its
eigenvalues.  $\bm{\mcal{T}}^{(0)}(k)$ is non-hermitian and has
complex eigenvalues for general values of $k$.  Denoting the largest
eigenvalue of this matrix by $\lambda^{(0)}_1$,we solve 
Eq.~\ref{eqT0eval1} by tuning $k$ until $\im{\lambda^{(0)}_1 (k=k_1)} 
= 0$; this determines the threshold $D_{th}^{(1)} = 1/ \lambda^{(0)}_1 (k=k_1)$. $k_1$ is the lasing
frequency of the first mode at threshold; the corresponding
eigenvector of $\bm{\mcal{T}}^{(0)}(k)$ gives the projection of the lasing mode
onto the CF states, $\bm{a}^1$, at threshold. The ``length" of
$\bm{a}^1$ is not determined from Eq.~\ref{eqT0eval1} but rises 
continuously from zero at threshold and is determined by the non-linear equation~(2)
infinitesimally above threshold.  The other, smaller eigenvalues of
$\bm{\mcal{T}}^{(0)}(k)$ define the non-interacting thresholds for
other modes with their frequencies determined by the same reality
condition, but the actual thresholds of all higher modes will differ
substantially from their non-interacting values due to the non-linear
term in Eq.~2 which now comes into play. The actual lasing
frequencies of higher modes have only a weak dependence on $D_0$ and
differ little from their non-interacting values (see Fig. 2, inset).

\begin{figure}
\centering
\includegraphics[width=30.0pc]{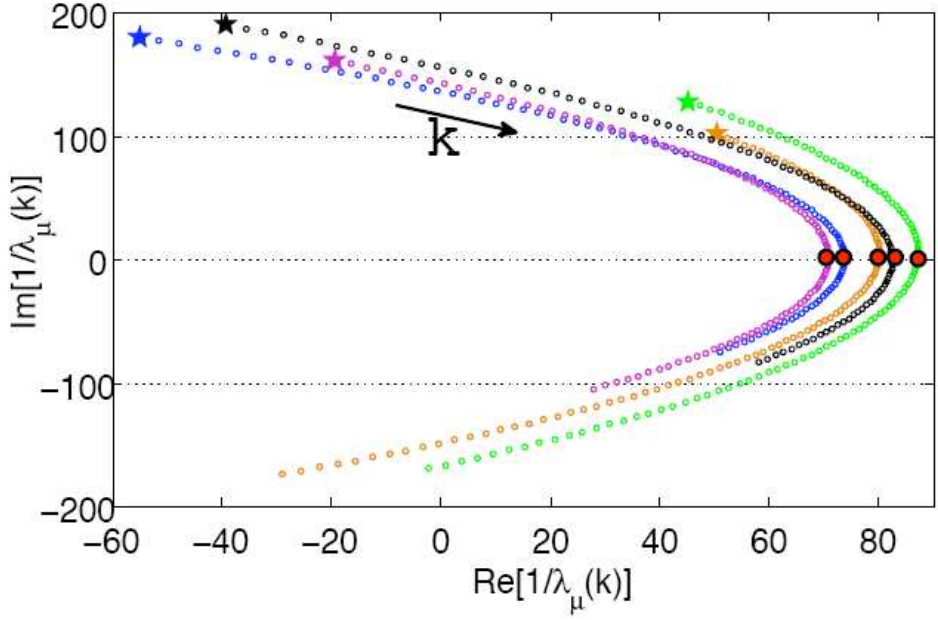}
\caption{Eigenvalue flow for the non-interacting threshold matrix.
Colored circles represent the trajectories in the complex plane of
$1/\lambda_\mu(k)$ as a function of $k$ for a scan
$k=(k_a-\gamma_\perp,k_a+\gamma_\perp)$ ($k_aR=30,\gamma_\perp R=1$).
Shown are only a few eigenvalues out of $N_{CF}=16$ modes included.
The direction of flow as $k$ is increased is indicated by the arrow.
Star-shapes mark the initial values of  $1/\lambda_\mu(k)$ at $k=k_a
- \gamma_\perp$. The full red circles mark the values at which the
trajectories intersect the real axis, each at a different value of
$k=k_\mu$, providing the non-interacting thresholds $D_{th}^{(\mu)}$.}
\label{fig:suplambdaflow}

\end{figure}

Above the threshold $D_{th}^{(1)}$ we solve the non-linear Eq.~2 by
an iterative method to be described below.  Assuming we have this
solution in hand at each value of $D_0$ we can construct the
generalized interacting threshold matrix
\begin{equation}
\mcal{T}_{mn}^{(1)}(k;D_0) = \frac{i\g}{\g - i (k-k_a)} \frac{1}{2k_a
(k - k_m(k))} \int  d\bxp \frac{ (1 + d_0(\bxp))\bar{\vp}^*_m (\bxp,k)
\vp_n (\bxp,k)}{\epsilon(\bxp)
(1 +  \Gamma (k_1) |\Psi_1 (\bxp)|^2)}.
\label{eqT1matrix}
\end{equation}
Here, the $D_0$ dependence of $\mcal{T}_{mn}^{(1)}$ derives from the
non-linear dependence of $\bm{a}^\mu$ (assumed to be determined by
the procedure described further below) on $D_0$. We will alternatively
use the notation $\mcal{T}_{mn}^{(1)}(k;\bm{a}^\mu)$ in what follows.
If we vary $k$ at fixed $D_0$ this linear operator will have a real
eigenvalue $\lambda^{(1)}_1(k_1) = 1/D_0$ reflecting the existence of
the first lasing mode.  To find the threshold for the second lasing
mode we vary $k$ until its second largest eigenvalue satisfies
$\im{\lambda^{(1)}_2 (k=k_2)} = 0$.  Similarly to the non-interacting
case, $k_2$ is the lasing frequency of the second mode and its
expected threshold is $D_{th}^{(2)} = 1/\lambda^{(1)}_2 (k=k_2)$.
This procedure generalizes in the obvious manner to the third and
higher thresholds.  Our algorithm for calculating the multimode
lasing states continuously monitors the relevant threshold matrix as
$D_0$ is varied to determine at each pump power how many lasing modes
are turned on.
The eigenvalues of these threshold matrices at a fixed $D_0$ have a
smooth flow in the complex plane and we can always associate a particular eigenvalue at an arbitrary $k$ with a particular lasing mode $\mu$ (which may not be yet turned on).  Henceforth $\mu$ is assumed to be ordered according to the order of turn-on.  The eigenvalue flow of the non-interacting
threshold matrix is illustrated in  Fig.~\ref{fig:suplambdaflow}; 
note that the initial values of each of the eigenvalues are different, as shown by the
star-shaped markers.

\paragraph*{Non-linear Solver:  }
The interacting threshold matrix $\bm{\mcal{T}}^{(1)}(k;D_0)$ provides us with the
starting values of the lasing frequencies $k_\mu$ and the threshold
solution ${\bf a^\mu}$ (up to a
proportionality constant) infinitesimally above threshold for the
non-linear lasing equations (2).
Equation~2 appears to be convenient for iterative solution, but it
must be further constrained
in order for this procedure to work.  $T_{mn}(k;\bm{a}^\mu)$ is
invariant under global phase rotations
$\bm{a}^\mu \rightarrow \ex{i\phi} \bm{a}^\mu$ and (2) only has a
unique solution when this
overall phase is fixed (the ``gauge" is fixed).  This phase can be
fixed in a trial solution, but it will be changed by each iteration
of the equation and so the correct procedure is to allow the trial
lasing frequency to flow under iteration so as to maintain the
desired global phase of the trial solution. In this manner the
interacting lasing frequencies can be found above threshold (note that these frequencies are different from the non-interacting or threshold values).
In practice, we choose the gauge by setting
$\im{a_{M_\mu}^\mu}=0$, where $M_\mu$ is the largest CF component of
the eigenvector $\bm{a}^\mu$ of the non-interacting threshold matrix
$\bm{\mcal{T}}^{(0)}(k)$.

\begin{figure}
\centering
\includegraphics[width=35.0pc]{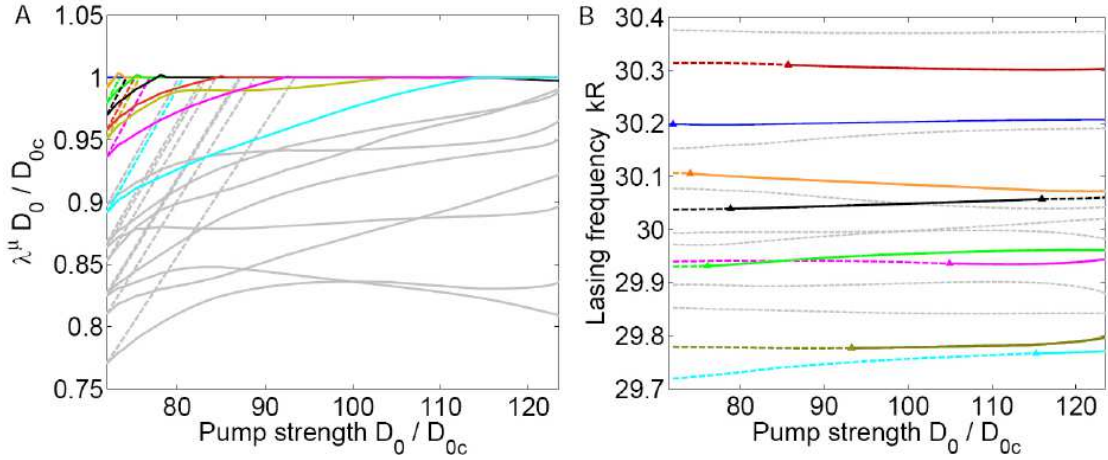}
\caption{Evolution of the
thresholds and lasing frequencies as a function of $D_0$. (A)
Evolution of the quantities $\lambda_\mu^{(i)} D_0(i)$ as the pump is
increased. $\lambda_\mu^{(i)}$ denotes the real $\lambda_\mu$
calculated at step $i$ of the discretization of the full $D_0$-range.
At a given step $i$, all modes $\mu$ which
are below the threshold (delineated by the line $\lambda_\mu D_0=1$)
are non-lasing at pump $D_0(i)$. Once a mode starts to lase its
corresponding value $\lambda_\mu D_0$ is clamped at $\lambda_\mu
D_0=1$. The full colored lines represent the modes which start lasing
within the calculated range of $D_0$. The dashed lines represent the
evolution $\lambda_\mu^{(0)} D_0(i)$ i.e. the evolution, had mode-mode
interactions not existed, while gray lines indicate modes which do
not turn on in the pump range shown. The color-coding is
identical to that of Fig.~2. (B) Analogous evolution of lasing
frequencies $k_\mu$ at which $\lambda_\mu^{(i)}(k)$ become real. The
non-lasing modes are drawn in dashed lines, turning into full lines
as the modes begin to lase. Gray lines represent modes which never
lase in the calculated range of $D_0$.  Note the absence of frequency
repulsion of non-lasing modes.}
\label{fig:supthreshevol}
\end{figure}

With this important modification the solutions to Eq.~2 are found
by increasing $D_0$ in small steps above the first threshold and
solving for the fixed point(s) of the equation, the vectors ${\bf
a^\mu}$, by iteration.  For this we use a multi-dimensional root
finder based on the Powell hybrid method. Convergence depends on the
quality of the initial approximation which in turn depends on how
fine the pump range is discretized. We find that close to the
thresholds the rate of convergence is in general slower as would be
expected for non-linear systems close to a bifurcation. At each value
of $D_0$ the associated interacting threshold matrix is constructed
from the non-zero ${\bf a^\mu}$ which have been found and monitored
to check if the next lasing mode has reached threshold and should be
included in the non-linear system. Note that the number of lasing
modes is not a monotonically increasing function of $D_0$; we find
that lasing modes can ``turn-off" due to strong modal interactions,
as described below (black mode in Fig.~2 and Fig.~\ref{fig:supthreshevol}).

The eigenvalues of the interacting threshold matrices as a function
of $D_0$ are very interesting because they show the strong effects of
mode competition.  If we plot $D_0 \lambda_\mu^{(0)}$ vs. $D_0$ these
are just straight lines intersecting unity at the non-interacting
thresholds; the interacting eigenvalues will be sub-linear, leading
to much higher thresholds, and some will even be decreasing with
increasing $D_0$, indicating modes which are completely suppressed by
mode competition and might never turn on. This behavior is shown in
Fig.~\ref{fig:supthreshevol} below along with the behavior of the 
lasing frequencies vs. $D_0$. Strong mode-mode interactions mediated by gain-saturation can
be studied in detail in Fig.~\ref{fig:supthreshevol}. For instance 
we observe that the turn-on of the black mode is delayed from $D_0/D_{0c} \approx 72$ to $D_0/D_{0c}
\approx 78$ due to interactions mainly with the orange mode, which
turns on earlier.  From
Fig.~\ref{fig:supthreshevol}(B) we see that the frequencies of these 
two modes shift closer to each other, which increases the interaction and results finally in the
turn-off of the black mode at about $D_0/D_{0c} \approx 117$, as seen in
Fig.~2 and Fig.~\ref{fig:supthreshevol}(A). At this point we observe 
a kink in the
intensity of the orange mode.  A similar interaction takes place
between the green and purple modes as described in the main text.

\subsection*{Materials and Methods 2: Collective contribution to the
laser frequency}

Here we provide details leading to Eq.~3 of the main text;
in this section, we will measure all frequencies from the atomic
transition frequency $k_a$ to simplify the equations (e.g. $k_\mu -
k_a \rightarrow k_\mu$). The gauge-fixing condition
$\im{a^\mu_{M_\mu}}=0$ leads to the following equation for the
corresponding lasing frequency (we will set $M_\mu=1$)
\be
k_\mu =
\frac{\g(q^\mu_1 Re[A_1^\mu] - \kappa^\mu_1
Im[A_1^\mu])}{(\g +\kappa^\mu_1) Re[A_1^\mu] - (k_\mu -
q^\mu_1)Im[A_1^\mu]} =
\frac{\g(q^\mu_1   - \kappa^\mu_1
\sigma_\mu)}{(\g+\kappa^\mu_1) - (k_\mu -
q^\mu_1)\sigma_\mu}.
\label{eqsuppkmuex1}
\ee
Here $k_1^\mu = q_1^\mu - i \kappa_1^\mu$
( $\kappa_1^\mu > 0$) is the CF
frequency of the largest contributing CF component,
\be
A_1^\mu \equiv \sum_{n=1}^{N_{CF}} a^\mu_n \int  d\bxp \frac{ (1 +
d_0(\bxp))\bar{\vp}^{\mu*}_1 (\bxp,k)
\vp_n^\mu (\bxp,k)}{\epsilon(\bxp) (1 + \sum_\nu \Gamma_\nu
|\Psi_\nu (\bxp)|^2)}.
\ee
and $\sigma_\mu \equiv Im[A_1^\mu] / Re[A_1^\mu]$.
Equation~\ref{eqsuppkmuex1} is exact, but it is useful to make 
minor approximations in order
to get a more easily interpreted result.   In our parameter range $\sigma_\mu $ is typically much less than one and so we can replace $k_\mu$ in the denominator of Eq.~\ref{eqsuppkmuex1} with its first 
order approximation (i.e. the result for $\sigma_\mu =0$), which is $k^{(0)}_\mu = \frac{q_1^\mu}{1+\kappa_1^\mu/\gamma_\perp}$.  This leads to the result of Eq.~3
\be
k_\mu \approx k^{(0)}_\mu [1 -
\tan[\phi^\mu_1] \frac{\kappa^\mu_1}{q^\mu_1}], \label{eqkmu2}
\ee
where $\tan[\phi^\mu_1] = Im[A_1^\mu]/Re[A_1^\mu]$ and we identify the
collective contribution to the lasing frequencies by
$k^{(c)}_\mu =  -k^{(0)}_\mu \tan[\phi^\mu_1] \kappa^\mu_1/q^\mu_1$.
The first term, $k^{(0)}_\mu$, is well-known from single mode lasing,
it represents the pulling of the cavity frequency $q^\mu_1$ towards
the atomic line ($k_a=0$ in our current convention). In the
fractional finesse limit of a DRL $\kappa_1^\mu  > \g$, it alone
would give $k_\mu \approx q_1^\mu (\g/\kappa_1^\mu) < q_1^\mu$, i.e.
a very large pulling of the lasing frequencies towards the atomic
line center.  This effect is seen in Fig.~1.  However the second term
in Eq.~\ref{eqkmu2} is the collective effect due to all the other CF states.
This contribution has no analog in conventional lasers and is random
in sign, as can be seen in Fig.~1, where some frequencies are pushed
towards and others away from the atomic line center due to this term.
The size of this effect depends on the magnitude of $\sigma_\mu \equiv
Im[A_1^\mu] / Re[A_1^\mu]$.  As noted, in our parameter range this
quantity is small; this is due to a remnant of the biorthogonality
relation.  Analysis of the quantity $A_1^\mu$ related to $T^\mu_{mn}$
suggests that at larger values of $k_aR$ and with more lasing modes
(higher above threshold) the quantity
$\sigma_\mu$ can  be large and of arbitrary sign.  When this is true
the approximation leading to Eq.~\ref{eqkmu2} will not be valid, but from
Eq.~\ref{eqsuppkmuex1} we see that the lasing frequencies will be 
dominated by the
collective effects of all the CF poles and will not be associated
with any single CF state or passive cavity resonance. The results
discussed in Materials and Methods~4 below support this conjecture.

\subsection*{Materials and Methods 3: Numerical calculation of CF
modes of the DRL and parameters}
\label{sect:supnumcf}

In our model for the DRL the region of uniform gain is assumed to be
a disk of radius $R$, on which the differential equation
\ref{eqcfgenev} is discretized with a polar mesh
($\rho(i)$,$\phi(j)$), $i=1,\ldots,N_\rho$, $j=1,\ldots,N_\phi$
chosen to be finer than the wavelength of the light $\lambda =
2\pi/k$ ($k$ is eventually to be set to the respective lasing
frequencies $k_\mu$). The resulting eigenvalue equation is
$\mcal{L}_{ij} \, \varphi_m(i,j) = k_m^2(k) \varphi_m(i,j)$
where $\mcal{L}_{ij} = -\frac{1}{\eps(\rho_i,\phi_j)}
\nabla^2_{\rho(i),\phi(j)}$, with $\nabla^2_{\rho(i),\phi(j)}$ the
discretized Laplacian in polar coordinates, see Ref.~({\it S2}). The static dielectric
disorder enters the discretized operator at each grid point
explicitly via $\eps(\rho_i,\phi_j)$ and can thus be chosen at will;
in what follows the dielectric function at each grid point takes the
values $\eps=\eps_1=(1.2)^2$ or $\eps=\eps_0=1$ randomly with roughly
20\% coverage
of ``nanoparticles". The outgoing CF boundary condition is imposed
by continuously connecting the solution of Eq.~\ref{eqcfgenev} and its
derivative to a superposition of outgoing Hankel functions,
$H_m^{(+)}(kR)e^{\pm i m \phi}$, on the boundary of the
gain-disk. The resulting boundary conditions are $k$-dependent and
can be written as
\begin{equation}
\varphi(N_\rho+1,\phi_j) = \sum_m a_m[\varphi(N_\rho)] \left( 1 +
k\delta \frac{H_m^{(+)\prime} (kR)}{H_m^{(+)} (kR)} \right) e^{i m
\phi_j}. \label{eqBC1}
\end{equation}
where $\delta = \rho_{i+1}-\rho_i$ and $a_m[\varphi(N_\rho)]$ are
the discrete angular Fourier coefficients of the solution on the
last ring $\varphi(N_\rho,\phi_i)$
\begin{equation}
a_m[\varphi(N_\rho)] = \frac{1}{2\pi} \sum_i^{N_\phi} \varphi
(N_\rho,\phi_i) \ex{-i m \phi_i}\label{eqBC2}
\end{equation}
This leads to a finite non-hermitian eigenvalue problem depending
parametrically on $k$ which is solved by customized linear algebra
packages for sparse matrices. Eqs.~\ref{eqBC1} and \ref{eqBC2} are
incorporated by adding a $N_\phi \times N_\phi$ block to the
discretized Laplacian $\mcal{L}_{ij}$ which renders it $k$-dependent.
The resulting solutions $\{ k_m(k), \varphi_m (\bx, k) \}$ are used
during the iteration procedure in the construction of the non-linear
operators $T_{mn}(k;\bm{a}^\mu)$.

The rest of the parameters used in the calculations presented are as
following: $\gamma_\perp R=1$, $k_a R=30$, $N_{CF}=16$.

\subsection*{Materials and Methods 4: Spatial structure of lasing modes}

\label{sect:supspatialstr}
The amplitudes $\bm{a}^\mu$ determine the spatial structure of the lasing modes via 
\[
\Psi_\mu (\bx) = \sum_{m=1}^{N_{CF}} a_m^\mu \varphi_m^\mu(\bx).
\]
Since these amplitudes evolve continuously from the relevant
eigenvectors of the
threshold matrices (see discussion above), it is useful to consider
their distribution from analysis of these matrices.  Construction of
the non-interacting threshold matrix
$\bm{\mcal{T}}^{(0)}$ (for uniform pumping) finds this matrix to be
close to diagonal and its eigenvectors to be localized on single CF
states, implying that at  the first threshold the lasing state is
primarily made up of one CF state.
The integral in Eq.~\ref{eqT0matrix} is not diagonal by 
biorthogonality due to
the factor $\epsilon (\bxp)$ in the denominator.  However, it can be
divided into an integral over the region with background dielectric
function $\eps_0$ and scattering centers with $\eps =\eps_1$.
Adding and subtracting appropriate quantities and using
biorthogonality one finds that
   $\mcal{T}_{mn}^{0} \propto \int d\bxp
\frac{\bar{\vp}^*_m (\bxp,k) \vp_n (\bxp,k)}{\epsilon(\bxp)} =
\frac{1}{\eps_0}\left[ \delta_{mn} - \frac{\eps_1 - \eps_0}{\eps_1}
\int_{g_1}  d\bxp
\bar{\vp}^*_m (\bxp,k) \vp_n (\bxp,k)\right]$, where $g_1$ denotes
the area of the gain region with $\eps = \eps_1$.  The second term
here is small due primarily to the fluctuating phases of the CF
states, but also due to the smallness of the prefactor and the fact
that we have taken  $g_1 \approx 20\%$ of the entire gain region.
When $\bm{\mcal{T}}^{(0)}$ is approximately diagonal, the quantity
$\sigma_\mu$ discussed above is small. As a result, at low pump powers
for the first lasing mode, $k_\mu \approx k_\mu^{(0)}$ with a small
collective contribution.

\begin{figure}
\centering
\includegraphics[width=30.0pc]{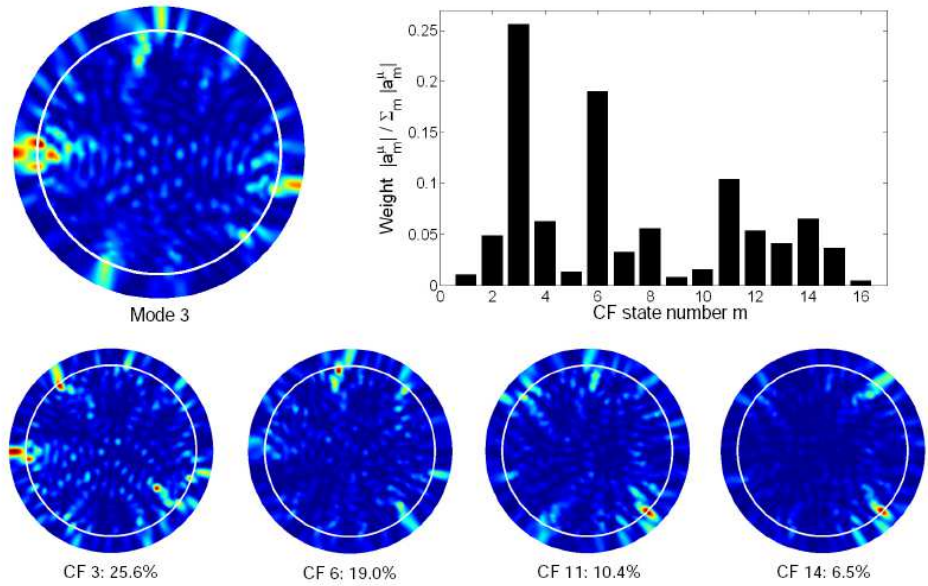}
\caption{Spatial structure of lasing modes. False color plot of the
spatial structure of the third mode (green curve in Fig.~2) and its expansion in the CF basis
(upper right) at $D_0/D_{0c} = 123.5035$. The structures of four of the CF states with largest weights are shown at the bottom. The maximum amplitudes 
of the field distributions in these plots are
rescaled to unity to facilitate comparison.}
\label{fig:supnonlwfs}
\end{figure}

However, if one analyzes the interacting threshold matrix
$\bm{\mcal{T}}^{(1)}(k;D_0) $ (and higher ones) one immediately sees
that the presence of lasing modes modifies the denominator in a
crucial manner: there is now a space-dependent term (the
``hole-burning term") which cannot be divided up into two constant
regions, but instead varies continuously throughout the gain medium.
As this term increases, the threshold matrices and the non-linear
operator $T^\mu_{mn}$ become more and more non-diagonal, and each
lasing mode becomes distributed  over many CF states.  Sufficiently
far above threshold, due to these interactions, the lasing modes
should lose all resemblance to any single CF state.  The modes that
turn on at higher pump powers tend to be more distributed over several CF states even
at threshold due to the less diagonal character of their higher
threshold matrices.  This is illustrated in 
Fig.~\ref{fig:supnonlwfs} which shows the
calculated decomposition in CF states at  $D_0/D_{0c} = 123.5035$ of the
green mode in Fig.~2.  For the mode chosen there are four CF states
with a weight greater than $10\%$. The 3rd CF state here has the
maximal contribution of $25.6\%$, while it is $55.8\%$ at
$D_0/D_{0c}=D^{1}_{th}=71.9519$.  The detailed behavior here merits further
study as to its dependence on size and type of disorder and on $k_aR$.

A second important observation about the spatial structure of the
lasing modes is that the amplitude of the lasing modes increase
quasi-exponentially towards the gain boundary, see
Fig.~4.  This is not due to high-Q passive cavity modes spatially
localized at the boundary (as it is for whispering gallery modes of
uniform spheres or cylinders); here there are no high-Q passive
cavity modes at all (see Fig.~1).  This effect is due to the high
gain needed to initiate lasing in such leaky systems.  Full
understanding of what determines the growth rate in this two-dimensional case
will also require further study.

\subsection*{Materials and Methods 5: ``Frequency repulsion" in DRLs}

We can now be more precise about the spatial correlation of modes
with nearly degenerate frequencies. By construction,
$\mcal{T}_{mn}(k)$ has a real eigenvalue $\lambda_\mu$ equal to
$1/D_0$ at the frequencies $k_\mu$ for each mode that is lasing. If
there were two modes with $k_\mu = k_\nu$, then the complex random
matrix $\mcal{T}_{mn}(k)$ would have an exact accidental
degeneracy. Such matrices are a set of measure zero in the ensemble,
and instead, as the eigenvalues $\lambda_\mu,\lambda_\nu$ approach each other in
the complex plane, there is an avoided crossing and strong mixing of
the eigenvectors $\bm{a}^\mu,\bm{a}^\nu$ (we see this mixing
numerically). Strong overlap of $\bm{a}^\mu,\bm{a}^\nu$ in turn implies
strong spatial correlation, strong hole-burning interaction and
suppression of the weaker intensity mode by the higher intensity
mode. The net effect is an apparent repulsion between lasing
frequencies, even though they are not themselves the eigenvalues of
a random matrix.  The crossing of lasing and non-lasing frequencies
(see Fig.~2) does not require degeneracy of the
threshold matrix and is allowed.

\clearpage

\clearpage

\clearpage

\clearpage

\subsection*{References and Notes}
\begin{itemize}
\item[S1.] 
Analytic formulas in Supporting Online Material have been simplified
by assuming $k_a$ is much greater than
all other frequencies measured from the atomic line center. This
approximation is not necessary and has not been used in the
calculations presented.
\item[S2.] S. Rotter, J.-Z. Tang, L. Wirtz, J. Trost, and J. Burgd\"orfer, Phys. Rev. B {\bf 62}, 1950 (2000).
\end{itemize}

\end{document}